\documentclass{ieeetj}
\usepackage{cite}
\usepackage{amsmath,amssymb,amsfonts}
\usepackage{graphicx}
\usepackage{textcomp}
\usepackage{xcolor}
\usepackage{booktabs}
\usepackage{array}
\usepackage{caption}
\usepackage{comment}
\usepackage{mathrsfs}
\usepackage[ruled,vlined]{algorithm2e}
\usepackage{float}
\usepackage{balance}
\usepackage{subcaption}
\usepackage{hyperref}
\hypersetup{hidelinks=true}
\def\BibTeX{{\rm B\kern-.05em{\sc i\kern-.025em b}\kern-.08em
    T\kern-.1667em\lower.7ex\hbox{E}\kern-.125emX}}
\AtBeginDocument{\definecolor{tmlcncolor}{cmyk}{0.93,0.59,0.15,0.02}\definecolor{NavyBlue}{RGB}{0,86,125}}

\def\OJlogo{\vspace{-4pt}$<$Society logo(s) and publication title will appear here.$>$}
\def\seclogo{\vspace{10pt}$<$Society logo(s) and publication title will appear here.$>$}

\def\OJlogo{}
\def\seclogo{}
\def\authorrefmark#1{\ensuremath{^{\textbf{#1}}}}

\begin{document}

\markboth{}{Bilican {et al.}}

\title{Content-Adaptive Inference for State-of-the-art Learned Video Compression}

\author{Ahmet Bilican\authorrefmark{1}, Student Member, IEEE, M. Akın Yılmaz\authorrefmark{2}, Member, IEEE,\\ and A. Murat Tekalp\authorrefmark{1}, Life Fellow, IEEE}
\affil{Dept. of Electrical \& Electronics Eng., Koç University, Türkiye}
\affil{ AI Research Team, Codeway Digital Services, Türkiye}
\corresp{Corresponding author: Ahmet Bilican (email: abilican21@ku.edu.tr).}
\authornote{``A. M. Tekalp acknowledges support from Turkish Academy of Sciences (TÜBA).''}

\begin{abstract}
While the BD-rate performance of recent learned video codec models in both low-delay and random-access modes exceed that of respective modes of traditional codecs on average over common benchmarks, the performance improvements for individual videos with complex/large motions is much smaller compared to scenes with simple motion. This is related to the inability of a learned encoder model to generalize to motion vector ranges that have not been seen in the training set, which causes loss of performance in both coding of flow fields as well as frame prediction and coding. As a remedy, we propose a generic (model-agnostic) framework to control the scale of motion vectors in a scene during inference (encoding) to approximately match the range of motion vectors in the test and training videos by adaptively downsampling frames. This results in down-scaled motion vectors enabling: i) better flow estimation; hence, frame prediction and ii) more efficient flow compression.
We~show that the proposed framework for content-adaptive inference improves the BD-rate performance of already state-of-the-art low-delay video codec DCVC-FM by up to 41\% on individual videos without any model fine tuning. We~present ablation studies to show measures of motion and scene complexity can be used to predict the~effectiveness of the~proposed framework. The code is available at \href{https://github.com/KUIS-AI-Tekalp-Research-Group/video-compression/tree/master/OJSP2025}{https://github.com/KUIS-AI-Tekalp-Research-Group/video-compression/tree/master/OJSP2025}.

\vspace{-3pt}
\end{abstract}



\maketitle

\section{INTRODUCTION}
\label{sec:intro}

\IEEEPARstart{N}{otable} progress has been achieved in learning-based video compression for both sequential P-frame (low-delay mode) and bi-directional B-frame (random access mode) codec models. The performance of video codecs is generally measured by BD-PSNR and BD-rate improvements over an anchor method, which are averaged over a test benchmark dataset. 
When the performance of learned codecs is investigated on a per sequence basis rather than on average, one can observe that the improvements for individual videos with complex/large motions are significantly smaller compared to videos with simple uniform motion. This is caused by the~inability of a single learned model to generalize in
the~presence of large motions; resulting in a loss of performance in both frame prediction/coding and coding of large motion vectors. 
The number of videos with large motions in the training set of most compression models is relatively small resulting in the generalization problem.
As a result, the performance of video codec models drop when the 
magnitude of motion vectors in test video are out-of-distribution compared to that seen in the training set. 

\begin{figure*}
    \centering
    \includegraphics[width=0.9\textwidth]{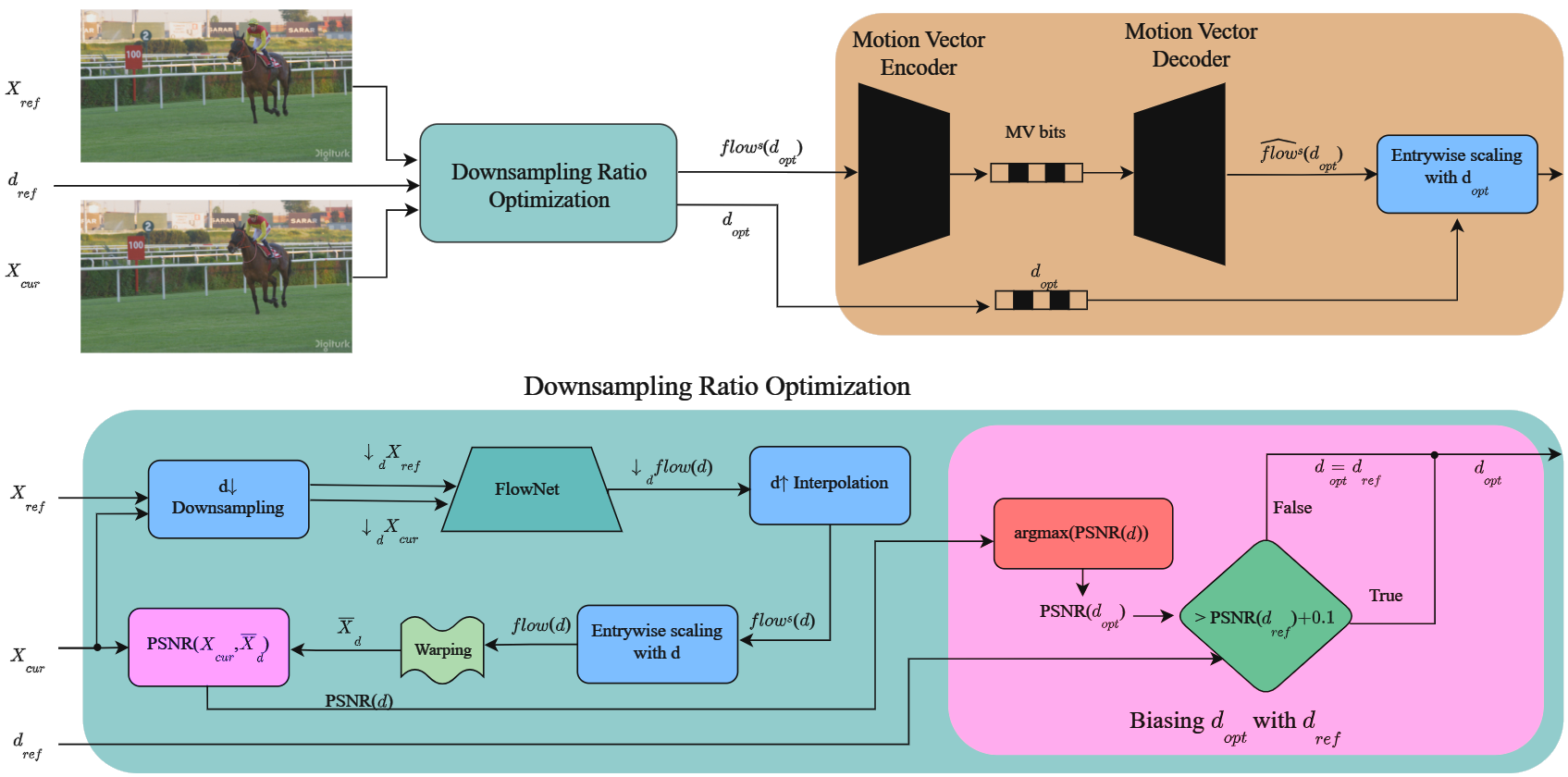} 
    \caption{The proposed content-adaptive inference framework. $d_{ref}$ denotes downsampling factor for previous frame.}
    \label{fig:overview}
\end{figure*}

One way to address this problem would be to supplement the training dataset of video compression models with many more videos with large and complex motions. However, this will require larger compression models to effectively learn the variety of different motion patterns in videos pushing learned video compression further away from practical usability in consumer devices.

In this paper, we propose an alternative simple, generic (model-agnostic) framework to control the scale of motion vectors on
a per frame basis to approximately match the range of motion vectors in the test and training videos during inference (encoding). Controlling the scale of motion vectors is an essential step (analogous to normalization), which has been generally ignored in the literature, for all learned video codecs to help improve both flow estimation (hence, frame prediction) accuracy as well as compression efficiency of flow vectors.
The temporal variability of motion vector magnitudes 
can be mitigated by adaptive frame
downsampling (only for flow estimation purposes) to control flow estimation accuracy and by scaling of motion vectors to allow for more efficient flow compression. 
After a brief discussion of related works in Section~\ref{sec:related}, we propose a generic framework for content-adaptive inference in Section~\ref{sec:method}. We demonstrate that our adaptive inference framework further improves the BD-rate performance of the state-of-the-art learned video codec DCVC-FM~\cite{dcvcfm} by up to 41\% on individual videos in Section~\ref{sec:results}. Section~\ref{sec:conc} concludes the paper.

\medskip

\noindent

\begin{figure*}[t!]
  \centering
  \begin{minipage}{0.3\linewidth}
    \includegraphics[width=\linewidth]{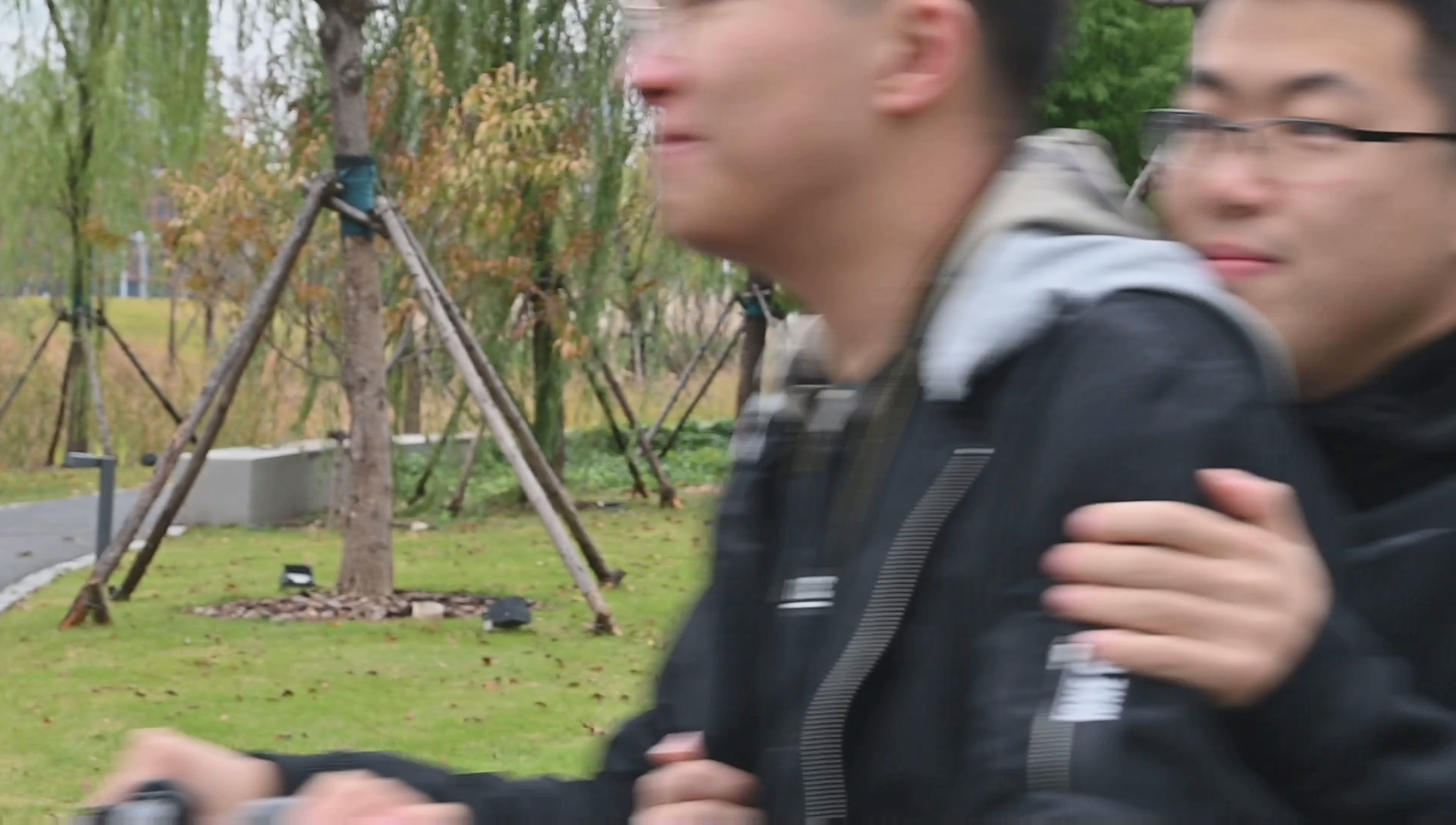} \vspace{-15pt} \\
    \centerline{a) current frame}
  \end{minipage}
  \begin{minipage}{0.3\linewidth}
    \includegraphics[width=\linewidth]{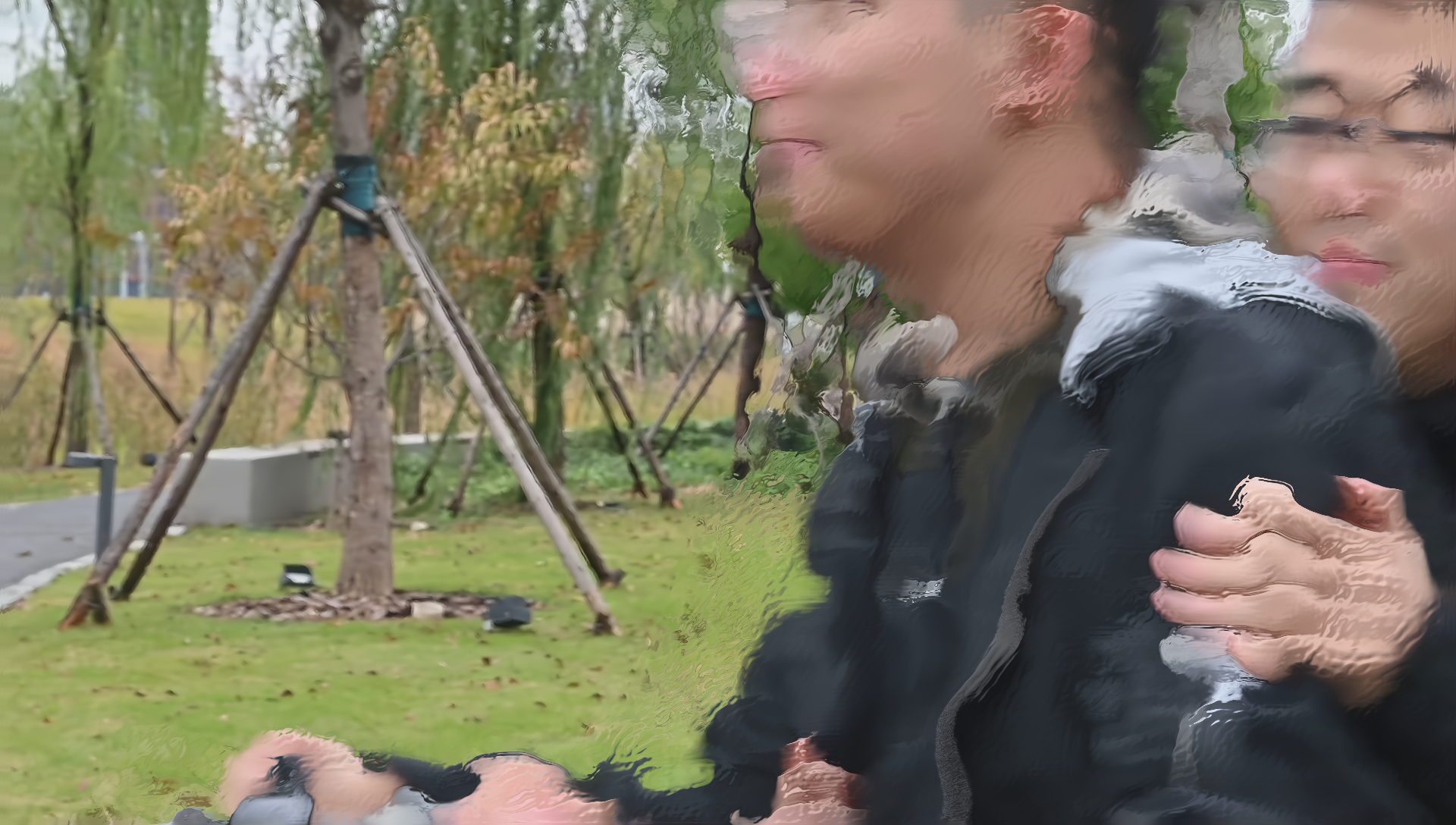} \vspace{-15pt} \\
    \centerline{b) warped frame w.o. optimization}
  \end{minipage}
  \begin{minipage}{0.3\linewidth}
    \includegraphics[width=\linewidth]{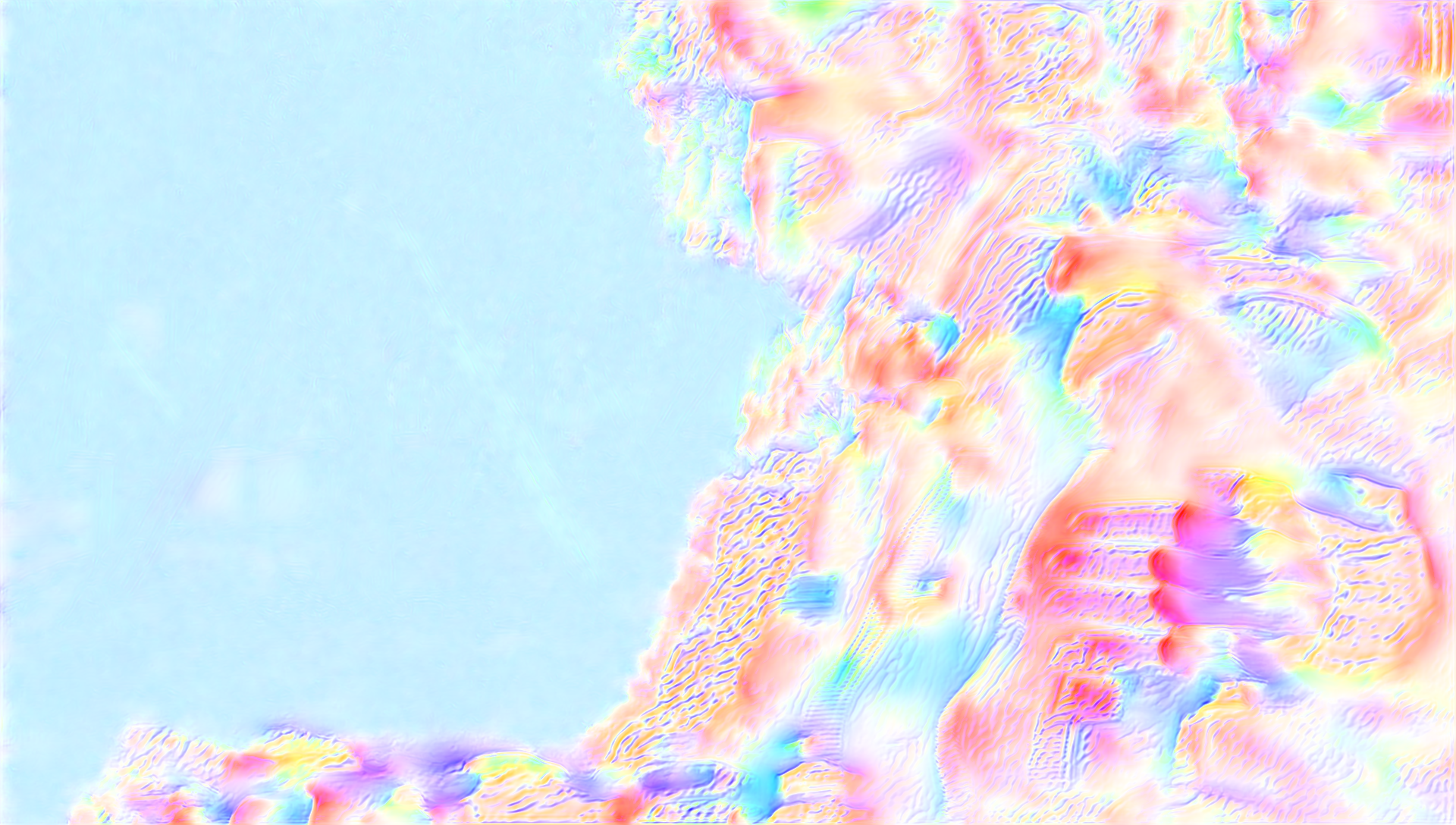} \vspace{-15pt} \\
    \centerline{c) original flow field}
  \end{minipage}
  
\vspace{2pt}

   \begin{minipage}{0.3\linewidth}
    \includegraphics[width=\linewidth]{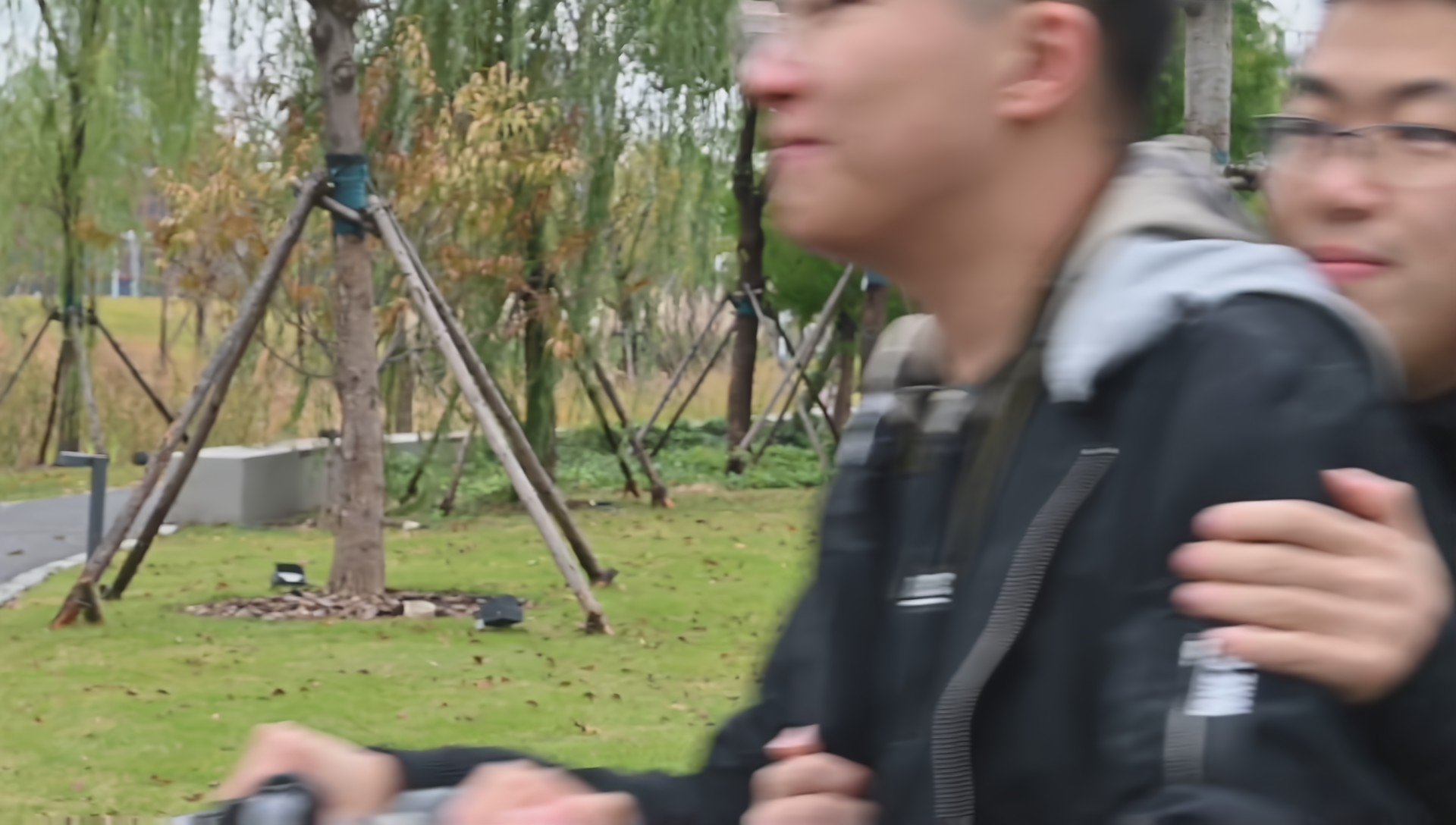} \vspace{-15pt} \\
    \centerline{d) reference frame}
  \end{minipage}
  \begin{minipage}{0.3\linewidth}
    \includegraphics[width=\linewidth]{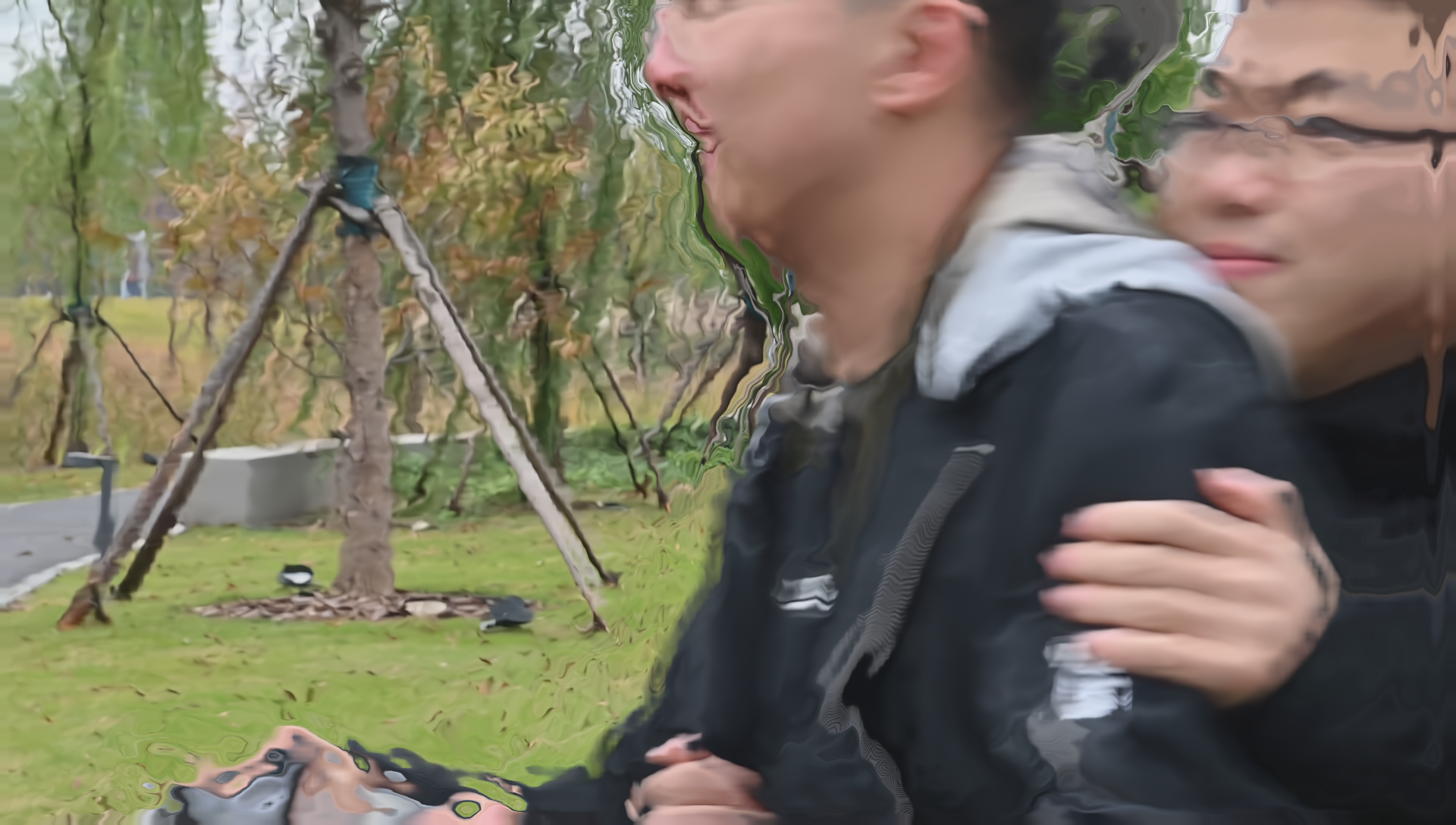} \vspace{-15pt} \\
    \centerline{e) warped frame after optimization}
  \end{minipage}
  \begin{minipage}{0.3\linewidth}
    \includegraphics[width=\linewidth]{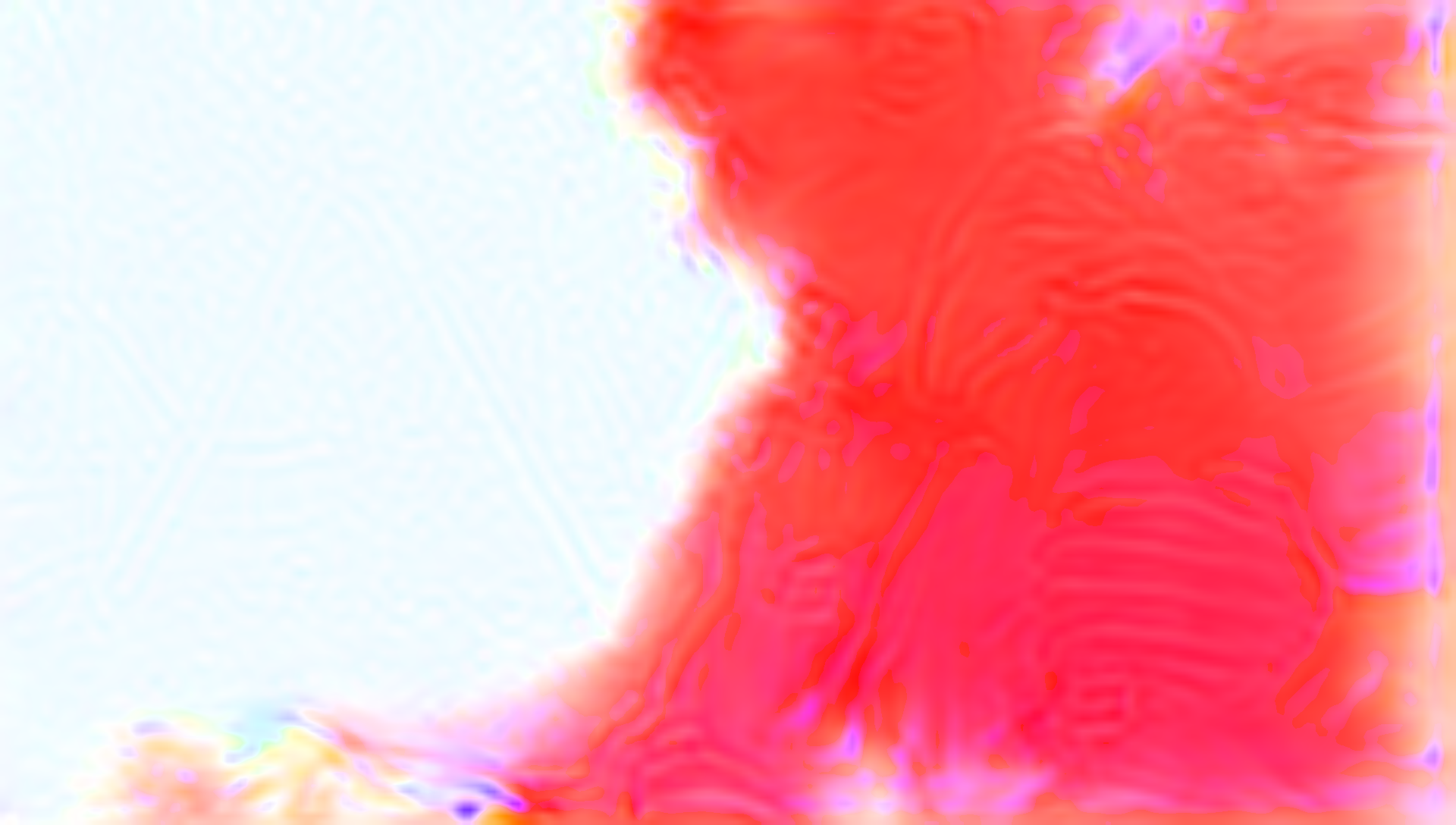} \vspace{-15pt} \\
    \centerline{f) optimized flow field}
  \end{minipage}

\vspace{1pt}  
\caption{The \textit{Bicycle\_Driving} sequence features two people riding a bicycle rapidly against a static background. Due to the high motion in the scene, the original model produces an inaccurate flow field, resulting in noticeable distortions in the warped frame. In contrast, the proposed method successfully captures the actual motion of the subjects, yielding a more precise reconstruction. Specifically, the warped frame without optimization (b) achieves a PSNR of 21.04 dB, while the optimized version improves the PSNR to 27.87 dB.}
\vspace{-6pt}
\end{figure*}

\section{Related Work and Novelty}
\label{sec:related}

We first review recent learned  unidirectional (low-delay mode) and bi-directional (random access mode) video compression models that would benefit from the proposed adaptive inference framework. We then discuss related recent test-time adaptive frame downsampling methods and how our proposed framework differs from them.





\subsection{Low-Delay Video Compression Models}
Low-delay sequential P-frame compression models have emerged as the predominant focus of learned video compression research. The pioneering work DVC~\cite{dvc} paved the way for end-to-end video codecs by jointly optimizing motion estimation, motion compression, and residual compression  under a single rate-distortion loss. Later, \cite{mlvc}~used multiple reference frames to generate more precise frame predictions and reduce residual information. The use of multiple references also enabled better motion vector prediction, reducing motion coding costs. Numerous other innovative ideas have been developed in entropy modeling~\cite{rlvc, mmvc} and model architectures~\cite{vct,scale_space,canfvc}.

More recent research elevated the coding process from pixel space to feature space~\cite{fvc}, while leveraging the advantages of conditional coding~\cite{conditional} to achieve significant improvements in the compression efficiency of learned codecs. In particular, the DCVC family of models has established state-of-the-art performance in sequential coding. In their initial DCVC model~\cite{dcvc}, the authors 
combined extracted feature map context with conditional coding and performed motion compensation in the feature space. This approach proved to be highly effective, establishing a new benchmark in the field. DCVC-TCM~\cite{dcvctcm} enhanced compression performance through the integration of multi-scale context modeling, representing another step forward in compression efficiency. In DCVC-HEM~\cite{dcvchem}, the authors introduced a sophisticated entropy model that effectively addresses both spatial and temporal redundancies. They propose a dual-prior architecture: a latent prior that exploits temporal correlations in latent representations, complemented by a parallel-friendly dual spatial prior for spatial redundancy reduction. Additionally, they implemented a content-adaptive quantization mechanism that operates at spatial-channel level, enabling both smooth rate adjustment within a single model and improved rate-distortion performance through dynamic bit allocation. DCVC-DC~\cite{dcvcdc} further advanced the field by expanding context diversity in both temporal and spatial dimensions. The~model introduced three key innovations: hierarchical quality patterns across frames for enriched long-term temporal contexts, group-based offset diversity with cross-group interaction for enhanced motion representation, and quadtree-based partitioning for improved spatial context diversity in parallel latent encoding. Most recently, DCVC-FM~\cite{dcvcfm} introduced the concept of feature modulation to conditional coding-based neural video compression. First, they expanded the supported quality range through latent feature modulation with learnable quantization scaling, employing a specially designed uniform quantization parameter sampling mechanism. Second, they tackled the quality degradation problem in long prediction chains by introducing a periodic feature refreshing mechanism for temporal feature modulation. The~model achieved substantial improvements in compression efficiency while reducing computational complexity, particularly under single intra-frame settings.


\subsection{Bi-directional Video Compression Models}
The pioneering work~\cite{yilmaz_icip20} introduced the~first end-to-end optimization framework for hierarchical bi-directional coding by accumulating the rate-distortion cost over fixed-size groups of pictures. The authors optimized hierarchical bi-directional flow estimation, flow compression, and frame prediction end-to-end all together using a single rate-distortion loss. HLVC~\cite{hlvc} proposed a three-layer hierarchical quality structure, combining bi-directional compression for high-quality frames with efficient single-motion estimation, enhanced by a weighted recurrent quality enhancement network at the decoder.  Several novel tools were proposed in \cite{yilmaz_tip21, yilmaz_icip22}, including learned masking, flow-field subsampling, and temporal flow vector prediction, further advancing the performance of learned hierarchical bi-directional compression. B-CANF~\cite{bcanf} exploited conditional augmented normalizing flows for B-frame coding. 
A novel two-layer conditional augmented normalization flows architecture was proposed~\cite{twocanf} that eliminated the need for motion information transmission. 
Instead of traditional motion coding, the method employed a low-resolution image compressor as a base layer, which combined with warped high-resolution images to generate conditioning signals for enhancement-layer coding. 
The recent work~\cite{yilmaz_icip23} improved motion compensation in hierarchical bi-directional compression through several innovations: a multi-scale deformable alignment scheme operating in feature space combined with multi-scale conditional coding, content-adaptive inference through online encoder updates, and a gain unit enabling flexible rate control across both intra-coded and bi-directionally coded frames. The framework further enhanced compression efficiency by incorporating a temporal latent extractor and parallel spatial-channel context model for improved entropy modeling. 
The key innovation in~\cite{yilmaz_icip24} was adapting the motion range during inference through adaptive frame downsampling based on both motion magnitude and hierarchy level, enabling a single flexible-rate model to effectively handle diverse motion scenarios. 
This adaptive approach provides state of the art performance in B-frame coding closing the gap between learned and traditional B-frame coding. Bi-directional DCVC~\cite{bidcvc} adapted the successful contextual modeling approach of DCVC to B-frame compression by introducing bi-directional motion difference context propagation, bi-directional contextual compression with temporal entropy modeling, and hierarchical quality structure-based training strategies, establishing a new benchmark in learned B-frame compression.

\subsection{Adaptive Frame Downsampling}
A critical challenge in learned video compression is the~domain shift in motion content of videos between training and test settings. This issue has only been recently recognized in the case of B-frame codec models, which are commonly trained with small GOP sizes due to dataset and memory limitations, while deploying them on larger GOPs during inference. The varying distances between past and future references across different hierarchy levels in hierarchical B-frames cause motion ranges to differ significantly from training conditions, particularly affecting complex scenes with large motions.
Two recent independent concurrent works addressed this challenge through different adaptive downsampling strategies for B-frame coding. The~first work, online motion resolution adaptation (OMRA)~\cite{omra}, determines optimal downsampling ratios through per-frame rate-distortion optimization before motion estimation, followed by motion compensation at the selected resolution. The~second work~\cite{yilmaz_icip24} proposed a prediction quality-driven approach for downsampling ratio selection, utilizing flow fields estimated at different resolutions to guide deformable convolution operations for more effective motion handling.
Subsequent work~\cite{fastomra} focused on improving downsampling factor selection through lightweight classifiers and proposed two implementations. The first implementation utilized a binary classifier trained with Focal Loss to choose between high and low-resolution motion estimation. The second employed a multi-class classifier trained with novel soft labels that incorporated rate-distortion cost knowledge. 
Note that all related works on adaptive frame downsampling are tailored to specific implementations of B-picture coding. None proposes a generic (model-agnostic) framework to address generalization of learned P-picture codecs to handle videos with out-of-distribution motion content.


\begin{figure}[t!]
    \centering
    \includegraphics[width=0.81\linewidth]{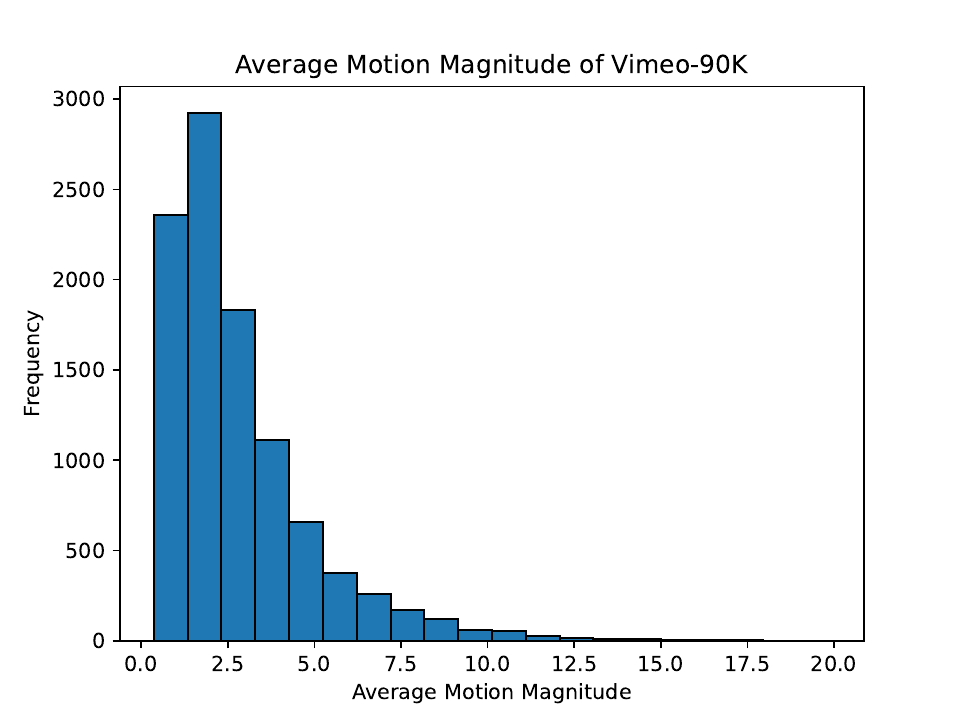} \vspace{-6pt}
    \caption{Distribution of average motion magnitude for 10,000 randomly sampled sequences from the Vimeo dataset.}
    \label{fig:vimeo}
\end{figure}

\subsection{Novelty}
This paper addresses an important problem in learned video compression: how can we improve the generalization ability of learned video codecs to videos that contain motions that are out-of-distribution compared to the training set.
This problem is exemplified by the weak performance of the state-of-the-art learned video codec DCVC-FM in videos with high motion, such as UVG\_Jockey, USTC\_BicycleDriving, USTC\_Snooker.

We show that this generalization problem can be addressed by controlling the scale of motion vectors in a scene during inference (encoding) 
by adaptively downsampling frames. The downsampling ratio per frame is selected such that the down-scaled motion vectors match the distribution of motion in the training set enabling: i) better flow estimation; hence, frame prediction and ii) more efficient flow compression. The proposed adaptive inference is an essential step for enabling state-of-the-art learned codecs to generalize to videos with large (out-of-distribution) motion  efficiently without expensive model fine-tuning.

Although frame downsampling has recently been used in B-frame codecs, its vital utility in enabling generalization of P-frame encoder models to videos with out-of-distribution motion has not been shown. Prior work also does not demonstrate under what conditions adaptive frame downsampling is an effective and essential tool for learned video codecs, which we show through extensive ablations. More~specifically, we show up to 41 \% BD-rate gains on individual videos with high motion magnitudes and low scene texture complexity over the state of the art DCVC-FM model~\cite{dcvcfm} in Section~\ref{sec:results}.

\begin{figure}[t!]
    \centering
    \includegraphics[width=0.81\linewidth]{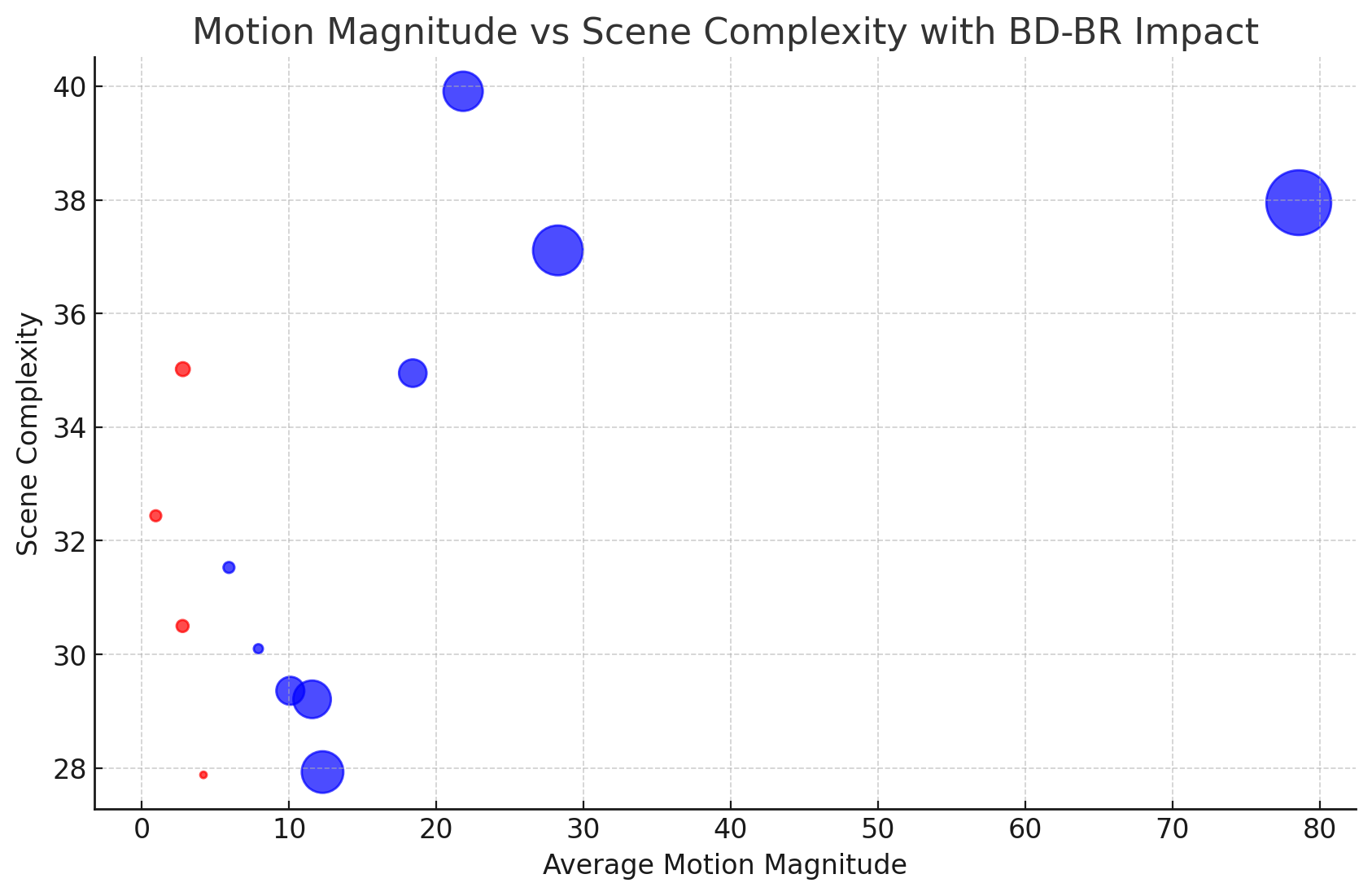} \vspace{-3pt}
    \caption{BD-rate improvements are depicted by the size of circles on scene complexity vs. average motion magnitude (performance predictors) plane. Blue shows a gain and red shows a loss. Threshold $T=5$ on motion magnitude was selected based on the plot.}
    \label{fig:bd-imp}
\end{figure}

\section{Content-Adaptive Inference in Learned Video Compression}
\label{sec:method}

We propose an adaptive inference framework that enables better generalization of a video compression model to larger displacements compared to those seen during training. The~key steps of the proposed framework are shown in Figure~\ref{fig:overview}. The operations of encoder and decoder modules are discussed in detail in Sections~\ref{sec_enc} and~\ref{sec_dec}, respectively.

\begin{algorithm}[!t]
    \caption{Adaptive Inference at the Encoder}
    \KwData{Current frame $\hat{X}_{cur}$, reference frame $\hat{X}_{ref}$ }
    \KwResult{Encoded flow field with scaling}
    \textbf{1. Select the best downsampling factor:} \\
    \quad Select the best downsampling factor $d_{opt}$ between current frame $\hat{X}_{cur}$, reference frame $\hat{X}_{ref}$ using Algorithm 2.

    \textbf{2. Estimate flow between downsampled frames:} \\
    \quad Estimate $\downarrow flow(d_{opt})$.

    \textbf{3. Upsample the flow field:} \\
    \quad Interpolate the motion field $\downarrow flow(d_{opt})$ to full frame resolution using bilinear interpolation to obtain $flow^{s}(d_{opt})$.

    \textbf{4. Determine whether to employ downsampling:} \\ 
       \quad \textbf{If} Average motion magnitude of $flow^{s}(d_{opt})\times d_{opt}$ is smaller than $T=5$  \textbf{then} $d_{opt}=1$
       
    \textbf{5. Encode the flow field:} \\
    \quad Encode scaled flow $flow^{s}(d_{opt})$ in full spatial resolution.

    \textbf{6. Send encoded flow field and $d_{\text{opt}}$ as side information:} \\
    \quad Send $d_{\text{opt}}$ to the decoder using 5 bits.

\end{algorithm}

\subsection{Adaptive Inference at the Encoder}
\label{sec_enc}

The key steps of the proposed framework, which are summarized in Algorithm 1, are: determining the optimal downsampling factor, flow estimation at reduced resolution, upsampling the flow field to full resolution, and finally encoding the full-resolution down-scaled flow field for transmission. These steps are discussed in detail below.

\subsubsection{Selection of the Downsampling Factor per Frame}
The core of our approach lies in determining the best downsampling factor out of a pre-defined set $\mathscr{D} = \{1,1.25, 1.5,\ldots,8.25, 8.5,8.75\}$ of candidate downsampling factors for each frame pair. Algorithm 2 outlines our selection process. For each candidate factor, we downsample both the current and reference frames, estimate the flow using the Flow Predictor network, and then up-sample and up-scale the resulting flow field. The selection criterion is the PSNR between the original current frame and its motion-compensated prediction. This process provides a trade-off between motion estimation accuracy and computational efficiency. \vspace{-10pt}

\subsubsection{Encoder Operation}
Once the best downsampling factor is selected, we estimate the flow field $\downarrow flow(d_{opt})$ between downsampled current and reference frames, producing reduced magnitude flow vectors at reduced resolution. This flow field is then spatially upsampled to the original frame resolution using bilinear interpolation, producing $flow^s(d_{opt})$. The interpolation of the flow field is necessary because DCVC-FM propagates motion field features to the next frame and encoding the motion field at different resolutions for each frame hinders the ability to propagate meaningful features. By performing the flow estimation at a lower resolution, our model can effectively handle larger displacements while maintaining computational efficiency. The full-resolution flow field is then encoded and sent along with the downsampling factor which requires only 5 bits of side information per frame.
\vspace{4pt}

\begin{algorithm}[!t]
    \caption{Select the Downsampling Factor per Frame}
    \KwData{Current frame $\hat{X}_{cur}$, reference frame $\hat{X}_{ref}$,   downsampling ratio $d_{ref}$ for the previous frame, a set of candidate downsampling factors $\mathscr{D}$}
    \KwResult{Best downsampling factor $d_{\text{opt}}$ for current frame}

    \For{$d \in \mathscr{D}$}{
        \textbf{1. Estimate flow} $flow(d)$ \\
        Downsample both $\hat{X}_{cur}$ and $\hat{X}_{ref}$ by a factor $d$, compute the flow with FlowPredictor, then upsample and scale the flow: \vspace{-4pt}
        \[
        \downarrow_d flow = \text{FlowPredictor}\left( \downarrow_d \hat{X}_{cur}, \downarrow_d \hat{X}_{ref} \right)
        \]

        \textbf{2. Upsample and scale the flow} \vspace{-4pt}
        \[
        flow^{s}(d) = \text{Interpolate}(\uparrow_d \text{flow, scale=d, bilinear)}
        \]
        \[
        flow(d) = d \cdot flow^{s}(d)
        \]

        \textbf{3. Compute the backwarped image} $\overline{X}_{cur}$ \vspace{-4pt}
        \[
        \overline{X}_{cur} = \text{backwarp}(\hat{X}_{ref}, flow(d))
        \]

        \textbf{4. Compute prediction PSNR} \vspace{-4pt}
        \[
        \text{PSNR}(d) = \text{PSNR}\left(X_{cur}, \overline{X}_{cur}\right)
        \]
    }
    \textbf{5. Choose} factor $d_{\text{opt}}$ that gives the best prediction PSNR: \vspace{-4pt}
    \[
    d_{\text{opt}} = \arg\max (\text{PSNR}(d))
    \]
    
    \textbf{6. Bias } $d_{\text{opt}}$ \textbf{ with } $d_{\text{ref}}$: \vspace{6pt}
    
    \quad \text{\textbf{If} } $\text{PSNR}(d_{\text{opt}}) < \text{PSNR}(d_{\text{ref}}) + 0.1$, \text{ then set }:
    \[
    d_{\text{opt}} = d_{\text{ref}}
    \]
\end{algorithm}

\subsection{Adaptive Inference at the Decoder}
\label{sec_dec}

The decoder operation follows the simple process outlined in Algorithm 3. Upon receiving the encoded flow field and the downsampling factor, it first decodes the flow field and then rescales it to the original scale using the received downsampling factor. This  ensures that the flow field is properly restored to its original scale while achieving better accuracy and better compression efficiency. The decoder method requires minimal modifications to the existing decoder model, as it only adds a scaling operation to the standard decoding pipeline.

\begin{figure}[t!]
  \centering
  \begin{minipage}{0.85\linewidth}
    \includegraphics[width=\linewidth]{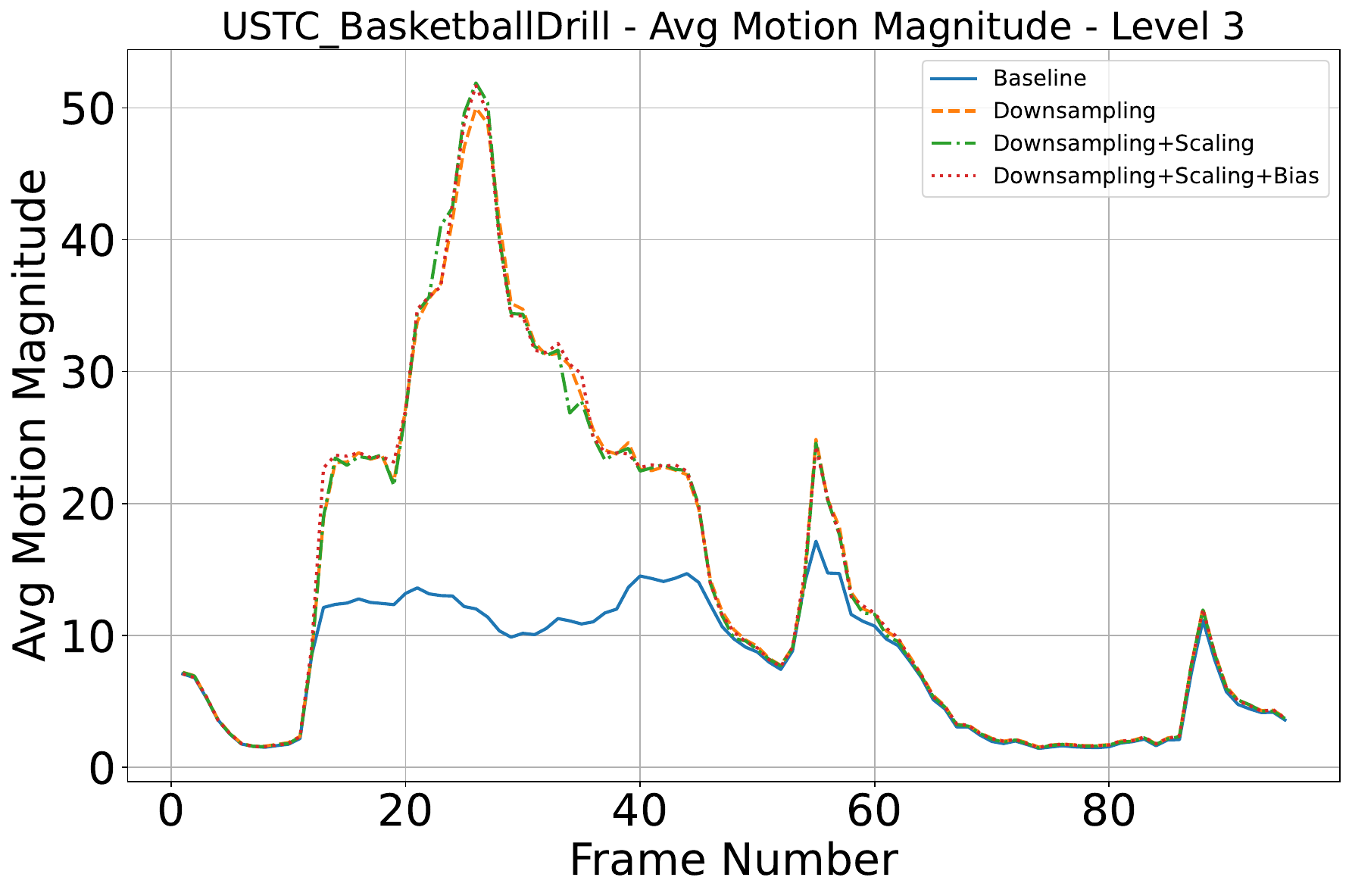} \vspace{-15pt} \\
    \centerline{(a)}
  \end{minipage}
 \vspace{3pt}
 
  \begin{minipage}{0.85\linewidth}
    \includegraphics[width=\linewidth]{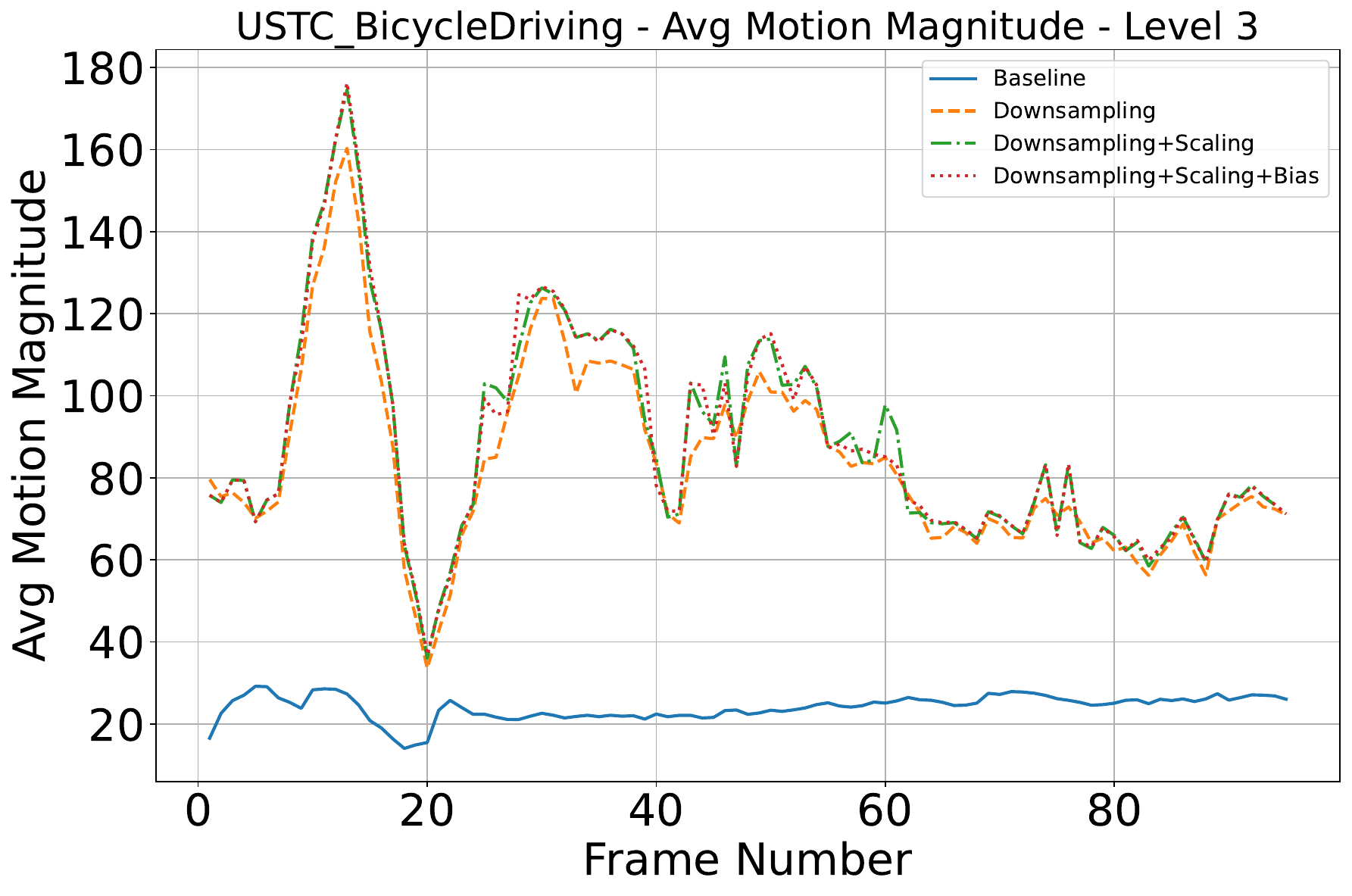} \\
    \centerline{(b)}
  \end{minipage} \vspace{-2pt}
    \caption{Comparison of the average motion magnitudes estimated with the corresponding techniques for a)~USTC\_BasketballDrill, b) USTC\_BicycleDriving. We see that the original estimation heavily underestimates the~amount of motion in a video because of its training data. 
    }
    \label{fig:motion-bias}
\end{figure}
\begin{algorithm}[!t]
    \caption{Adaptive Inference at the Decoder}
    \KwData{Downsampling ratio $d_{\text{opt}}$, compressed $flow^s(d_{opt})$}
    \KwResult{Decoded flow field at the original scale}

    \textbf{1. Decode flow field:} \\
    \quad Decode the flow field received from the encoder.
    
    \textbf{2. Rescale decoded flow field:} \\
    \quad Multiply each entry in the decoded flow field by $d_{\text{opt}}$ to obtain the flow field at the original scale.
\end{algorithm}

\section{Experimental Results}
\label{sec:results}

\begin{table*}[t]
\centering
\caption{BD-rate (RGB) comparison of proposed method vs. DCVC-FM over videos with different motion content} \vspace{-1pt}
\begin{tabular}[0.8\textwidth]{l|cc|c|cc|}
\cline{2-6}
\multicolumn{1}{c|}{} &
  \multicolumn{2}{c|}{\textbf{Sequence Statistics}} &
  \textbf{\% BD-BR} &
  \multicolumn{2}{c|}{\textbf{Ablations (\% BD-BR)}} \\ \hline
\multicolumn{1}{|c|}{\textbf{\begin{tabular}[c]{@{}c@{}}Sequence\\ Name\end{tabular}}} &
  \begin{tabular}[c]{@{}c@{}}Avg Motion\\ Magnitude\end{tabular} &
  \begin{tabular}[c]{@{}c@{}}Scene \\ Complexity \end{tabular} &
  \begin{tabular}[c]{@{}c@{}}Downsampling\\ +Scaling +Bias\end{tabular} &
  \begin{tabular}[c]{@{}c@{}}Downsampling\\ +Scaling\end{tabular} &
  Downsampling \\ \hline
\multicolumn{1}{|l|}{UVG\_Beauty}           & 2.78      & 35.02     & 0.00          & 0.00          & 0.00          \\ \hline
\multicolumn{1}{|l|}{UVG\_Bosphorus}        & 2.22      & 34.86     & 0.00          & 0.00          & 0.00          \\ \hline
\multicolumn{1}{|l|}{UVG\_HoneyBee}         & 0.69      & 35.54     & 0.00          & 0.00          & 0.00          \\ \hline
\multicolumn{1}{|l|}{UVG\_Jockey}           & 28.26     & 37.11     & -24.28          & -22.87          & -13.42          \\ \hline
\multicolumn{1}{|l|}{UVG\_ReadySetGo}       & 7.91                    & 30.10                   & -0.82           & 0.01           & -1.43           \\ \hline
\multicolumn{1}{|l|}{UVG\_ShakeNDry}        & 1.34                    & 33.60                   & 0.00           & 0.00           & 0.00           \\ \hline
\multicolumn{1}{|l|}{UVG\_YachtRide}        & 4.89                    & 30.63                   & 0.00           & 0.00           & 0.00           \\ \hline
\multicolumn{1}{|l|}{USTC\_Badminton}       & 5.91                    & 31.53                   & -1.18           & -1.20           & -0.52           \\ \hline
\multicolumn{1}{|l|}{USTC\_BasketballDrill} & 12.27                   & 27.93                   & -16.97           & -16.36           & -18.54           \\ \hline
\multicolumn{1}{|l|}{USTC\_BasketballPass}  & 11.56                   & 29.41                   & -13.85           & -13.41           & -13.46           \\ \hline
\multicolumn{1}{|l|}{USTC\_BicycleDriving}  & 78.59                   & 37.95                   & -41.13           & -39.78           & 18.61           \\ \hline
\multicolumn{1}{|l|}{USTC\_Dancing}         & 2.76                    & 30.50                   & 0.00           & 0.00           & 0.00           \\ \hline
\multicolumn{1}{|l|}{USTC\_FourPeople}      & 0.94                    & 32.44                   & 0.00           & 0.00           & 0.00           \\ \hline
\multicolumn{1}{|l|}{USTC\_ParkWalking}     & 18.40                   & 34.95                   & -7.44           & -7.40           & -7.52           \\ \hline
\multicolumn{1}{|l|}{USTC\_Running}         & 4.18                    & 27.88                   & 0.00           & 0.00           & 0.00           \\ \hline
\multicolumn{1}{|l|}{USTC\_ShakingHands}    & 10.08                   & 29.36                   & -7.68           & -8.02           & -9.38           \\ \hline
\multicolumn{1}{|l|}{USTC\_Snooker}         & 21.82                   & 39.91                   & -15.14           & -14.77           & 12.46           \\ \hline
\multicolumn{1}{|l|}{\textbf{}}             & \textbf{}               & \textbf{}               & \textbf{-7.56}  & \textbf{-7.28}  & \textbf{-1.95}  \\ \hline
\end{tabular}
\label{table:main}
\end{table*}

\begin{figure}[t!]
  \centering
  \begin{minipage}{0.85\linewidth}
    \includegraphics[width=\linewidth]{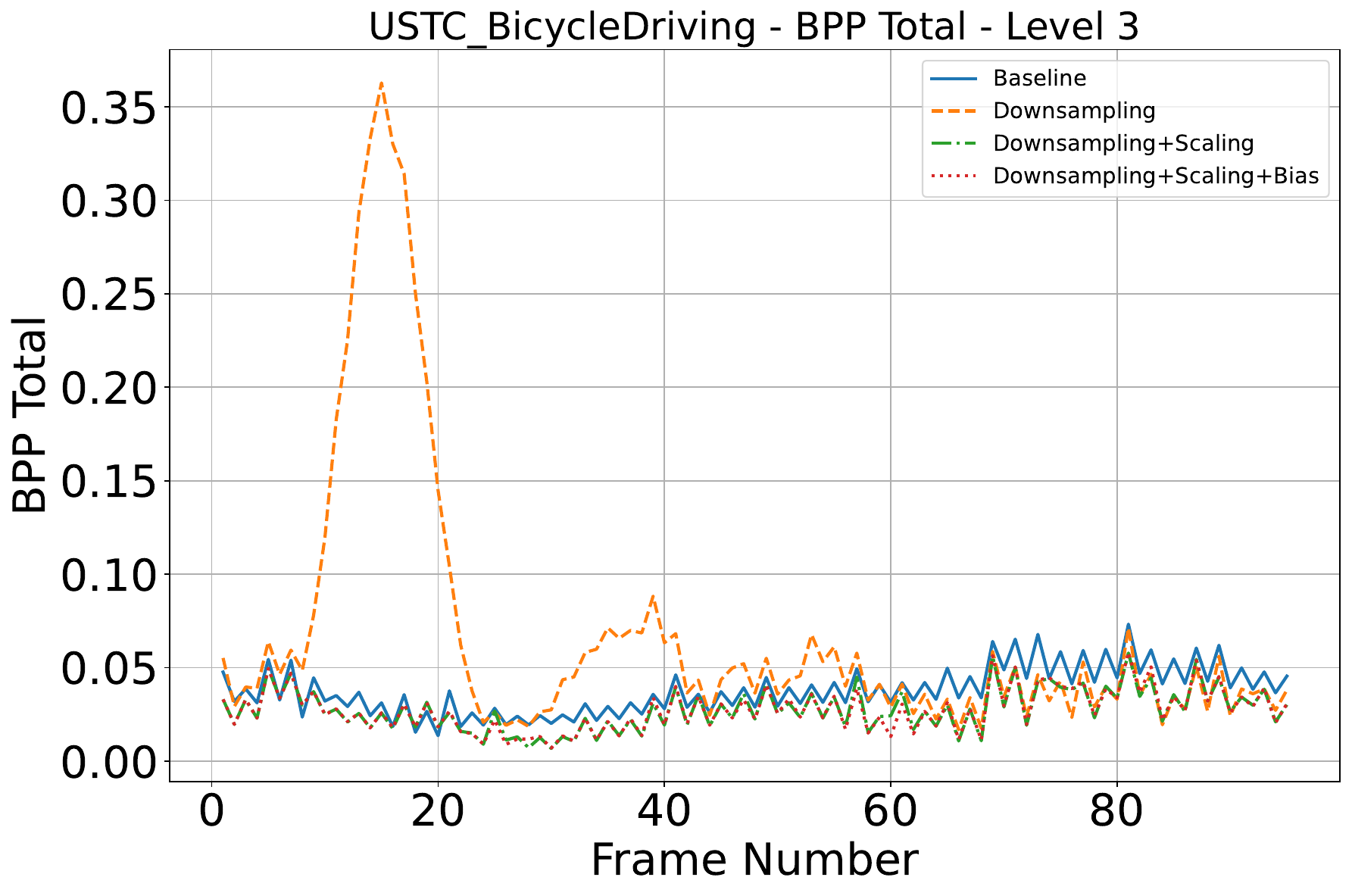} \vspace{-15pt} \\
    \centerline{(a)}
  \end{minipage}
 \vspace{3pt}
 
  \begin{minipage}{0.85\linewidth}
    \includegraphics[width=\linewidth]{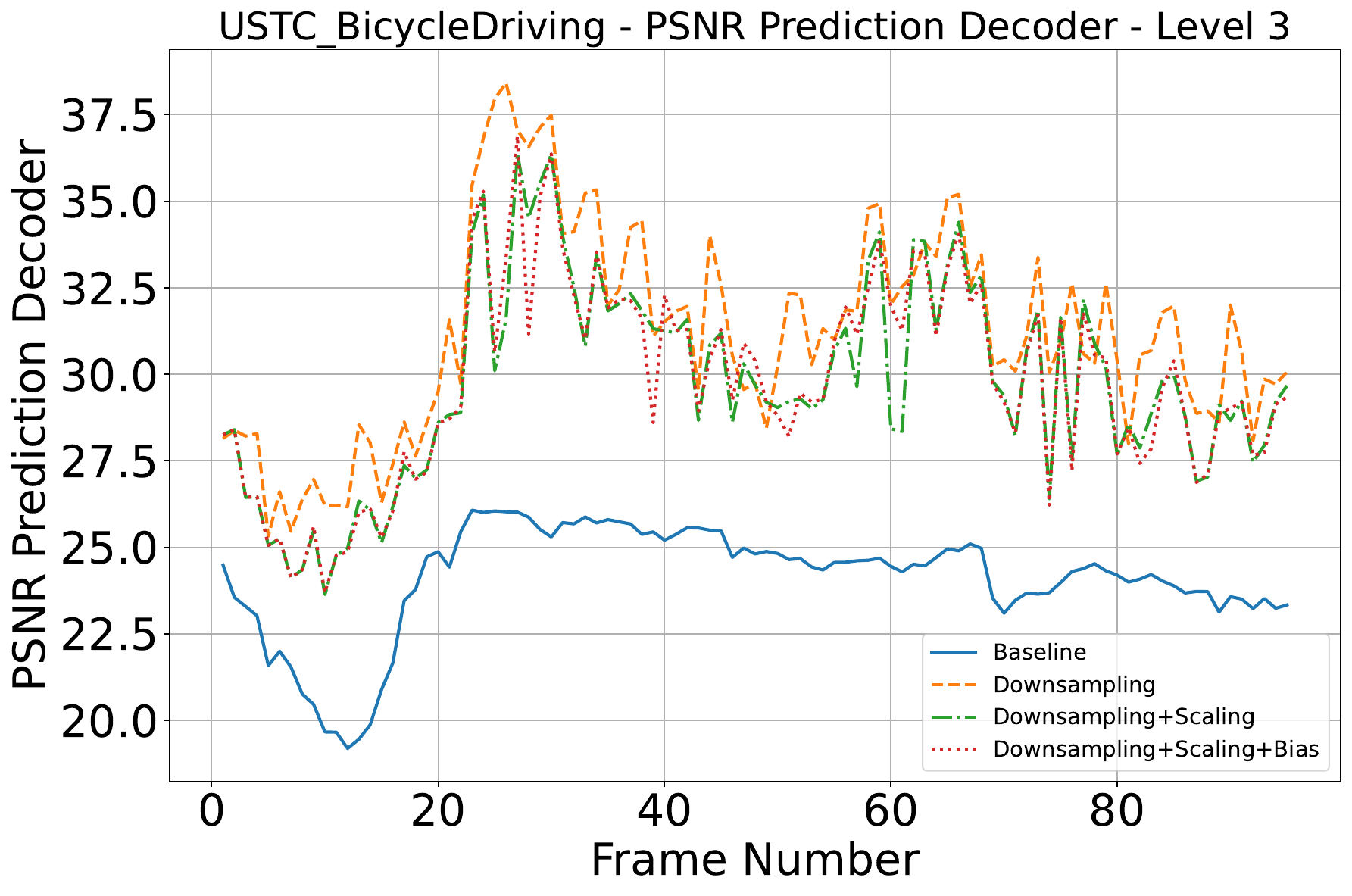} \\
    \centerline{(b)}
  \end{minipage} \vspace{-2pt}
  \caption{Analysis of rate (BPP) of (a) motion and (b)~prediction PSNR at decoder with and without scaling. Motion scaling allows efficient encoding of motion without noticeable loss of prediction PSNR.
  }
\label{fig:motion-analysis}
\end{figure}

In all experiments reported in this section, we modified the~DCVC-FM encoder and decoder for adaptive inference as described in Section~\ref{sec:method} without any retraining or fine-tuning of DCVC-FM model parameters.  We use GOP size -1 as in DCVC-FM config. We evaluate our framework on two widely-used benchmark datasets: UVG~\cite{uvg} and USTC~\cite{ustc}, which contain diverse video content with varying degrees of motion and scene texture complexity.

\subsection{Evaluation of BD-Rate Performance}
Our comprehensive experiments reveal that the proposed training-free adaptive inference framework via adaptive frame downsampling delivers exceptional results when the range of motion in test videos exceeds the motion range present in the training data and the scene texture complexity of the scene is not very high. In order to support this observation with quantitative evidence, we propose measures of motion and scene texture complexity in the following and then correlate the performance of our framework with these measures in Table \ref{table:main}.
\vspace{4pt}

\noindent \textbf{Average Motion Magnitude of Sequences} \\
We measure motion range by the average of magnitude of flow vectors in a video. To evaluate the motion range of the training set,
we randomly sampled 10,000 sequences from Vimeo and plotted a histogram of the average motion magnitudes in Figure \ref{fig:vimeo}. We see that the sequences in Vimeo mostly contain motion magnitudes around 2-5 pixels. Our findings show that while DCVC-FM performs admirably when the average motion magnitude of the test video aligns with the training distribution, when the motion magnitude is substantially higher, we observe poor optical flow estimation and poor compression efficiency, resulting in lower overall performance. Table~\ref{table:main} clearly shows that our framework effectively handles these out-of-distribution cases. We also see that the flow estimator heavily underestimates the motion present in high motion sequences in Figure \ref{fig:motion-bias}, and its effect on the prediction PSNR in Figure \ref{fig:motion-analysis}.
\vspace{4pt}

\noindent \textbf{The Calculation and Effect of Scene Texture Complexity} \\
In order to quantify scene texture complexity, we first downsample a frame by a factor of 4 with bicubic  antialiasing filter, then upsample it again using bicubic upsampling, and calculate the PSNR between the original and upsampled frames. If the frame has high scene texture complexity, the upsampled image will be blurred and the PSNR will be low. For frames with simpler texture details, the PSNR will be higher. Hence, lower PSNR indicates higher scene texture complexity. We can observe in Table~\ref{table:main} that our framework provides the best improvements when the motion magnitude is high and scene texture complexity is low.
\vspace{4pt}

Inspection of Table~\ref{table:main} reveals
our method achieves exceptional performance improvements on videos containing challenging high-motion and low scene texture complexity content. We demonstrate extraordinary BD-rate gains, -41.13\% improvement on USTC\_BicycleDriving sequence, -24.28 \% improvement on UVG\_Jockey sequence, -16.97 \% improvement on USTC\_BasketballDrill, and -15.14 \% improvement on USTC\_Snooker sequence. These impressive improvements over the DCVC-FM baseline clearly demonstrate that our adaptive inference framework allows DCVC-FM to generalize to these out-of-distribution videos.

When comparing our results with those reported in the USTC dataset paper~\cite{ustc}, we observe that our method significantly closes the performance gap between DCVC-FM and VVC in sequences where DCVC-FM traditionally struggles, while maintaining its superior performance in scenarios where it already excels. This balanced improvement leads to remarkable BD-rate gains of up to 41 \% compared to the state-of-the-art DCVC-FM, establishing a new benchmark in learned video compression.
\vspace{-4pt}


\subsection{Ablation Studies}

In this section, we first investigate the effect of upsampling the estimated motion vectors at the encoder (before motion coding) vs. at the decoder (sending the downsampling factor). We also analyse the benefit of biasing the downsampling factor towards the value at the previous frame.
\vspace{4pt}

\begin{figure*}[t!]
  \centering
  \begin{minipage}{0.3\linewidth}
    \includegraphics[width=\linewidth]{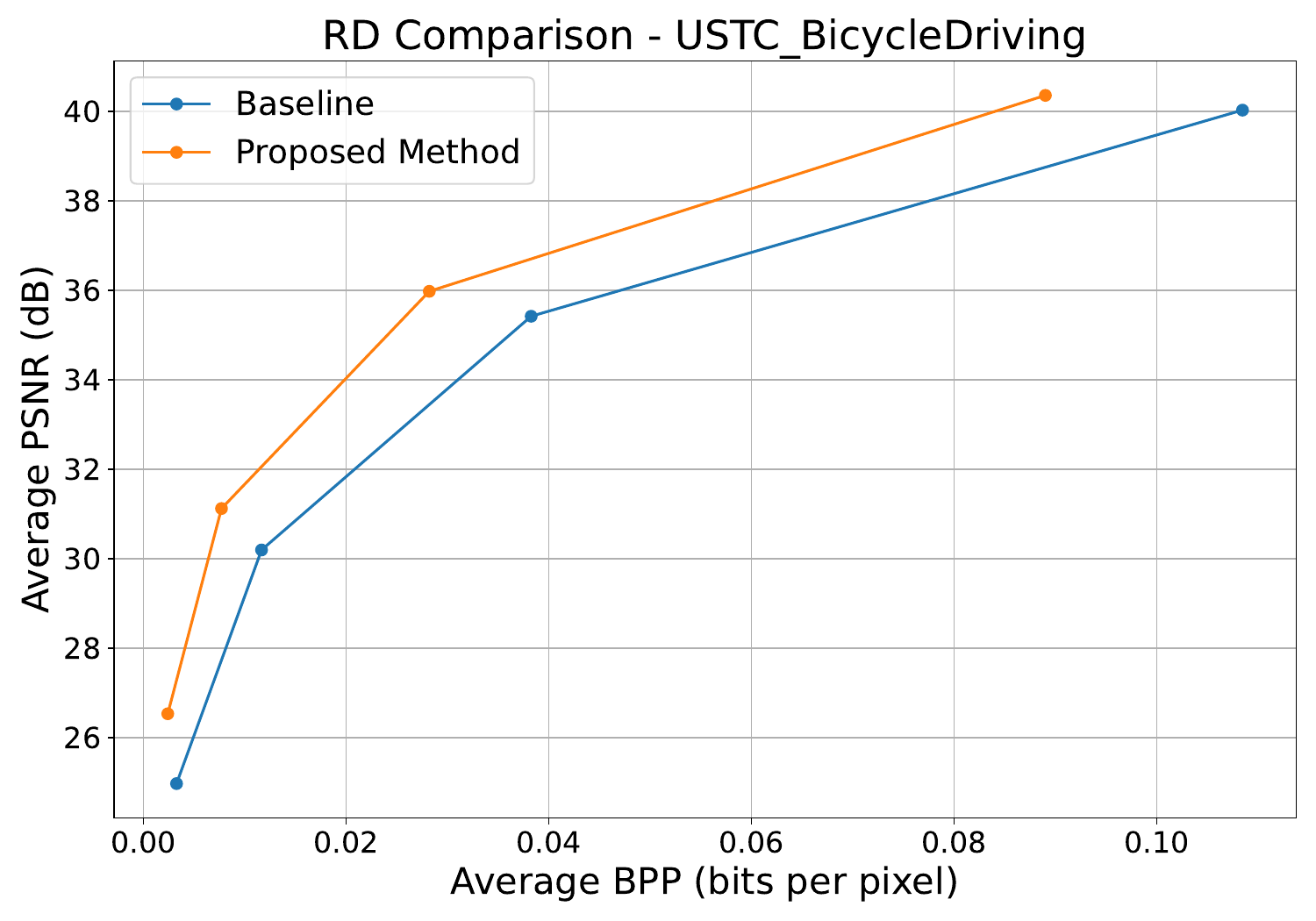} \vspace{-18pt} \\
    \centerline{(a)}
  \end{minipage}
  \begin{minipage}{0.3\linewidth}
    \includegraphics[width=\linewidth]{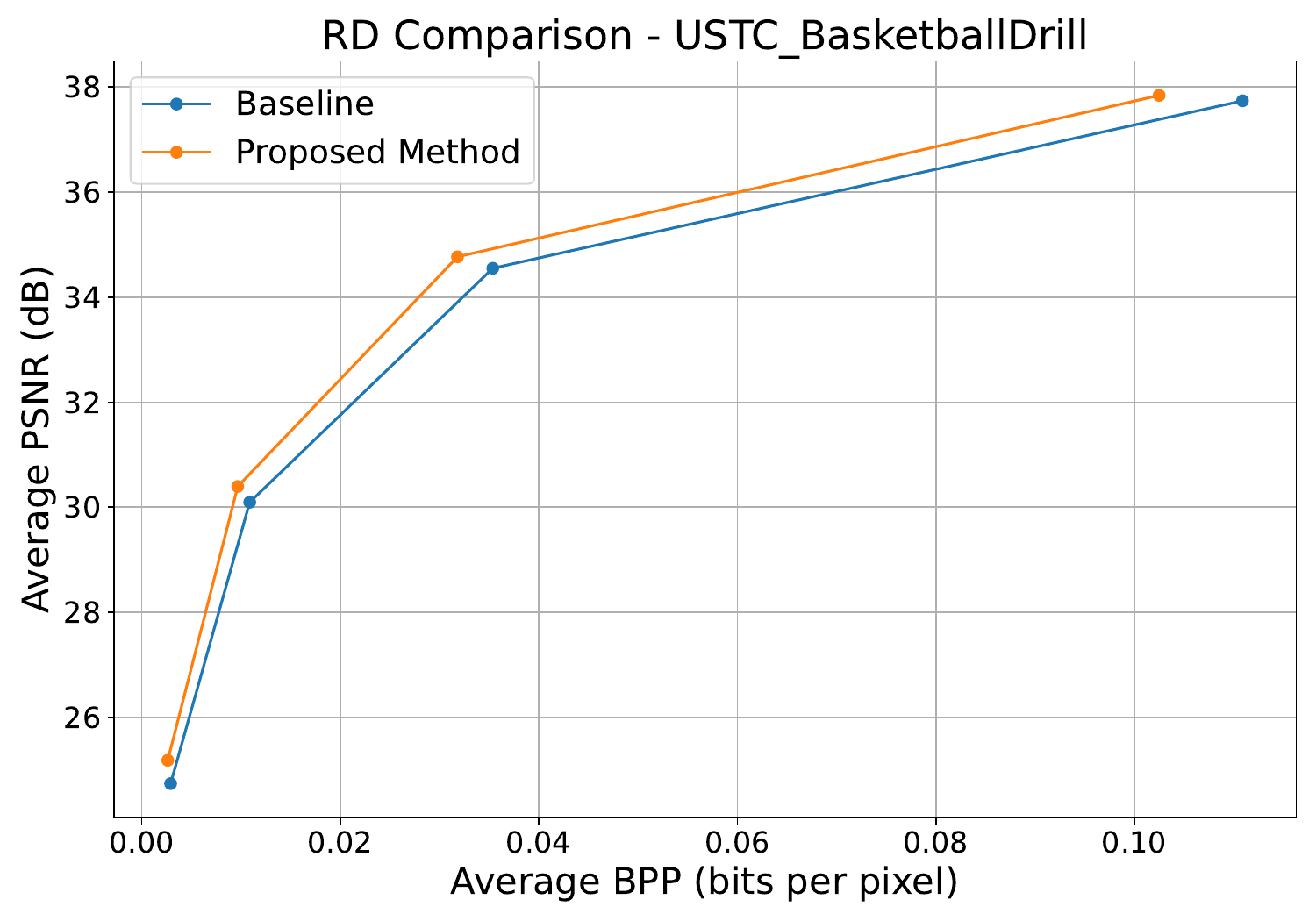} \vspace{-18pt} \\
    \centerline{(b)}
  \end{minipage}
  \begin{minipage}{0.3\linewidth}
    \includegraphics[width=\linewidth]{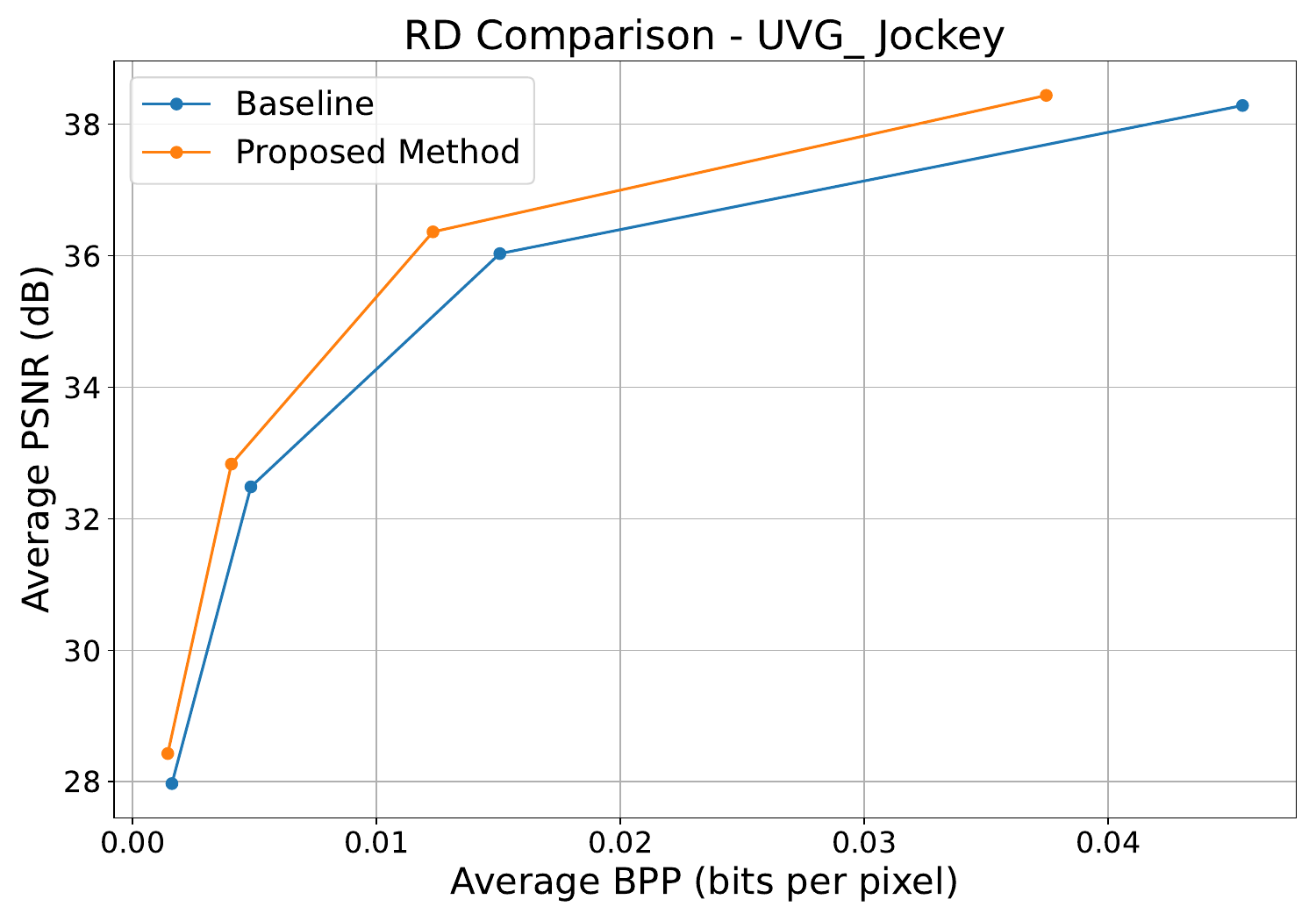} \vspace{-18pt} \\
    \centerline{(c)}
  \end{minipage}
  
\vspace{-1pt}  
\caption{RD curves (RGB) of baseline and proposed adaptive inference on some video sequences (a) USTC\_BicycleDriving, (b) USTC\_BasketballDrill, (c) UVG\_Jockey} \vspace{-6pt}
\label{fig:RD-Curves}
\end{figure*}

\noindent \textbf{Effectiveness of Motion Vector Scaling} \\
One of our key innovations lies in the handling of motion vector scaling.  Down-sampling motion vectors (without scaling) leads to an increase in motion bits when the magnitude of motion vectors is large as shown in Figure~\ref{fig:motion-analysis}a.
Scaling the magnitude of the motion vectors ensures that the~estimated motion field remains within distribution, which in turn reduces the total bits per pixel required with no considerable loss in the prediction PSNR at the decoder as shown in Figure~\ref{fig:motion-analysis}b.
On average, we observe that scaling the motion before compression results in a BD-rate improvement of $-5.12\%$, as shown in Table~\ref{table:main}. Notably, in the USTC\_BicycleDriving sequence —where DCVC-FM performs significantly worse than VVC — downsampling+scaling approach yields a substantial BD-rate reduction of $-58.39\%$ with respect to downsampling only.

\vspace{6pt}

\noindent \textbf{Benefits of Biasing} \\
Our framework incorporates an innovative biasing approach that proves particularly effective when prediction PSNRs between different downsampling ratios show minimal differences. When the difference between prediction PSNRs is minimal, it is preferable to retain the previous ratio, as this allows us to propagate more similar features from the previous motion field to assist in encoding the current one. When temporally consecutive motion fields are assigned different optimal downsampling ratios, the coding efficiency drops  because of the increase in discontinuity between temporally consecutive motion vectors, ultimately impairing coding efficiency. 
\vspace{4pt}

\noindent \textbf{Prediction PSNR as a Selection Criterion} \\
In selecting the downsampling ratio, we utilize prediction PSNR rather than rate-distortion loss to identify the best choice. This approach is driven by the fact that both the optical flow estimator and the motion compressor are trained on the same dataset. Our hypothesis is that when the flow estimator performs accurately—meaning it yields a better motion field prediction and consequently a high prediction PSNR—the resulting motion statistics will also be well-suited for the motion compressor.


\vspace{4pt}

\subsection{Computational Complexity}

The proposed adaptive inference method introduces an optimization step to select the optimal downsampling ratio before computing the final flow field. This selection process primarily utilizes the flow estimator and warping components from the original DCVC-FM model, rather than the entire encoder-decoder architecture.

We evaluated the computational overhead introduced by the proposed content-adaptive encoding. Experimental results, obtained using an NVIDIA RTX A6000 GPU and averaged over 10 frames across 4 different bit rates for the~USTC\_BicycleDriving sequence, show the following: \vspace{-2pt}
\begin{itemize}
\item Average time to fully encode one frame using the~baseline DCVC-FM encoder is approximately 941.6~ms.
\item Our optimization step adds an average overhead of 536.4 ms per frame. This corresponds to the time required to evaluate the adaptive downsampling ratio using the flow estimation and warping steps.
\item Hence, the total average encoding time per frame increases approximately a 57\% over the baseline.
\item The decoding time is not affected by the proposed adaptive frame downsampling at the encoder.
\end{itemize}

While this represents a quantifiable increase in encoding time, it allows the framework to achieve substantial BD-rate improvements, especially on challenging sequences, as demonstrated in Table \ref{table:main}.

Crucially, this overhead is significantly lower than the~alternative approach of performing a full rate-distortion~(RD) optimization for each potential downsampling ratio. If our method relied on evaluating the complete RD cost for each candidate ratio (assuming, for instance, 32~potential ratios were tested), it would necessitate running the entire encoder-decoder pipeline 32 times per frame. This~would dramatically increase computational complexity. Our approach avoids this prohibitive cost by using the~efficient heuristic that higher prediction PSNR correlates with better RD performance, a strategy supported by the~results in~\cite{yilmaz_icip24}, which indicates that such exhaustive RD optimization may not yield proportionally significant gains.

\section{Conclusion}
\label{sec:conc}

The ability of learned models to handle out of distribution data is crucial to generalize their performance. The~amount of motion in a video is a key parameter that affects the~performance of video processing and compression models per video. The number of high motion videos in the training set of state-of-the-art video compression models is relatively small; hence, high-motion videos are generally considered to be difficult to code.\\ This paper proposes two essential steps to improve the~performance of learned video compression models for high-motion videos: \\ i)~adaptive frame downsampling (for motion estimation only) enables better estimation of large motions by bringing the~range of motion vectors closer to those seen in the~training set; hence, allows for better frame prediction in the presence of large motions, \\ ii)~adaptive motion vector scaling per frame at the~encoder enables better compression of large motion vectors.\\
We demonstrate that the proposed training-free adaptive inference framework enables better generalization of already state of the art video compression models to videos with large motion. In particular, we show that we obtain up to 41\% better BD-rates on individual videos with large motion and relatively simple scene complexity and 7.56\% BD-rate improvement on average (UVG+USTC datasets) over the~state of the art DCVC-FM anchor.

\bibliography{main}
\bibliographystyle{IEEEtran}
\end{document}